\documentclass[a4paper]{jpconf}
\usepackage{graphicx}
\newcommand{\logg}{\mbox{$\log g$}}
\newcommand{\Teff}{\mbox{$T_\mathrm{eff}$}}

\newcommand{\msun}{\ensuremath{\, {\rm M}_\odot}}
 
\newcommand{\rsun}{\ensuremath{\,{\rm R}_\odot}} 
\newcommand{\ion}[2]{\mbox{#1\,{\sc #2}}}

\def\etal{{et\,al.}\ }
\def\sdss{SDSS\,1228+1040}
\begin{document}
\title{Spectral modeling of gaseous metal disks around DAZ white dwarfs}

\author{Klaus Werner, Thorsten Nagel, and Thomas Rauch}

\address{Institute for Astronomy and Astrophysics, Kepler Center for
  Astro and Particle Physics, University of T\"ubingen, Germany}

\ead{werner@astro.uni-tuebingen.de}

\begin{abstract}
We report on our attempt for the first non-LTE modeling of gaseous metal
disks around single DAZ white dwarfs recently discovered by G\"ansicke
\etal and thought to originate from a disrupted asteroid. We assume a
Keplerian rotating viscous disk ring composed of calcium and hydrogen
and compute the detailed vertical structure and emergent spectrum. We
find that the observed infrared \ion{Ca}{ii} emission triplet can be
modeled with a hydrogen-deficient gas ring located at $R$\,=\,1.2\,R$_\odot$,
inside of the tidal disruption radius, with \Teff$\approx$\,6000\,K and a
low surface mass density of $\approx$\,0.3\,g/cm$^{2}$. A disk having
this density and reaching from the central white dwarf out to
$R=1.2$\,R$_\odot$ would have a total mass of $7\cdot 10^{21}$\,g,
corresponding to an asteroid with $\approx$160\,km diameter.
\end{abstract}

\section{Introduction: Dust around DAZ white dwarfs}

More than two decades ago Zuckerman \& Becklin (1987) announced the
discovery of an IR excess around the DAZ white dwarf G29$-$38. The
white dwarf itself is enriched in metals. Considering the short
sedimentation timescales in the photosphere this implies that the star
is accreting matter at a relatively high rate (Koester \etal
1997). Since no cool companion has been found at G29$-$38, the hypothesis
was put forward that a dust cloud around the white dwarf causes the IR
excess. In fact, the presence of dust has been confirmed by
\emph{Spitzer} observations (Reach \etal 2005). Graham \etal (1990)
concluded that the dust is located in the equatorial plane. Subsequently, further
DAZ white dwarfs with potential dust disks were found (Becklin \etal
2005, Kilic \etal 2005, 2006). As a possible origin of these disks
tidally disrupted comets were discussed (Debes \& Sigurdsson 2002) and,
more likely because of the absence of H and He, disrupted asteroids
(Jura 2003).

\section{Gas disks around DAZ white dwarfs}

Recently, signatures of a gas disk were discovered in Sloan Digital Sky
Survey (SDSS) spectra of two DAZ white dwarfs (G\"ansicke \etal 2006,
2007). The spectra display double-peaked emission lines of the infrared
\ion{Ca}{ii} triplet $\lambda\lambda$\,8498, 8542, 8662\,\AA.

In the present paper, we concentrate on one of these two white dwarfs,
namely \sdss, because its emission line profiles are more prominent. 
The white dwarf's atmospheric parameters are \Teff\,=\,22\,020\,K and
\logg\,=\,8.24. The derived stellar mass $M_{\rm
WD}$\,=\,0.77\,\msun\ and radius $R_{\rm WD}$\,=\,0.011\,\rsun\ are quantities that
enter our disk model. The photospheric magnesium abundance is 0.8
times solar. The spectrum neither exhibits radial velocity variations nor
photometric variability. Besides the calcium emission lines only two other
weaker emission features are seen (\ion{Fe}{ii} $\lambda\lambda$\,5018,
5169\,\AA). In particular, hydrogen and helium emissions are not
discovered. It is concluded that the Ca and Fe emission lines stem from
a metal-rich Keplerian disk around a single white dwarf.  The
\ion{Ca}{ii} line profiles are double peaked emission lines with a
peak-to-peak separation of 630\,km/s, i.e., the Keplerian rotation
velocity is $v\,\sin\,i$\,=\,315\,km/s. There is a clear violet/red asymmetry
in the double-peaked profiles, well known from a similar phenomenon in
Be star disks that is ascribed to one-armed spiral waves.

From a spectral analysis with a kinematical LTE emission model
G\"ansicke \etal (2006) conclude that we see a geometrically thin,
optically thick disk at high inclination ($i$\,=\,70$^\circ$). The inner and
outer disk radii are $R_{\rm in}$\,=\,0.64\,\rsun\ and $R_{\rm
out}$\,=\,1.2\,\rsun, respectively. While the outer disk radius is quite sharply confined
because of the steep line wings, the value derived for the inner disk radius
possibly just marks the inner edge of the \ion{Ca}{ii} line emission
region and not the physical inner disk edge, i.e., $R_{\rm in}$ could
reach down to the white dwarf's surface. They also conclude that the gas
temperature in the disk is around 4500--5500\,K. Since the tidal
disruption radius for a rocky asteroid at the white dwarf is 
$R$\,$\approx$\,1.5\,\rsun, it is possible that the material in the gas disk may be a
disrupted asteroid whose dust was sublimated by the white dwarf's radiation
field. It seems that \sdss\ could be the hot counterpart to G29$-$38
and other cool DAZ stars harboring dust disks, but in the meantime DAZ
white dwarfs with a dust disk and as hot as \sdss\ were discovered
(e.g., Jura \etal 2007).

It is our aim to compute a more realistic model for the gas disk around
\sdss\ and we are reporting here very first provisional results.

\begin{figure}
\begin{center}
\includegraphics[width=11cm]{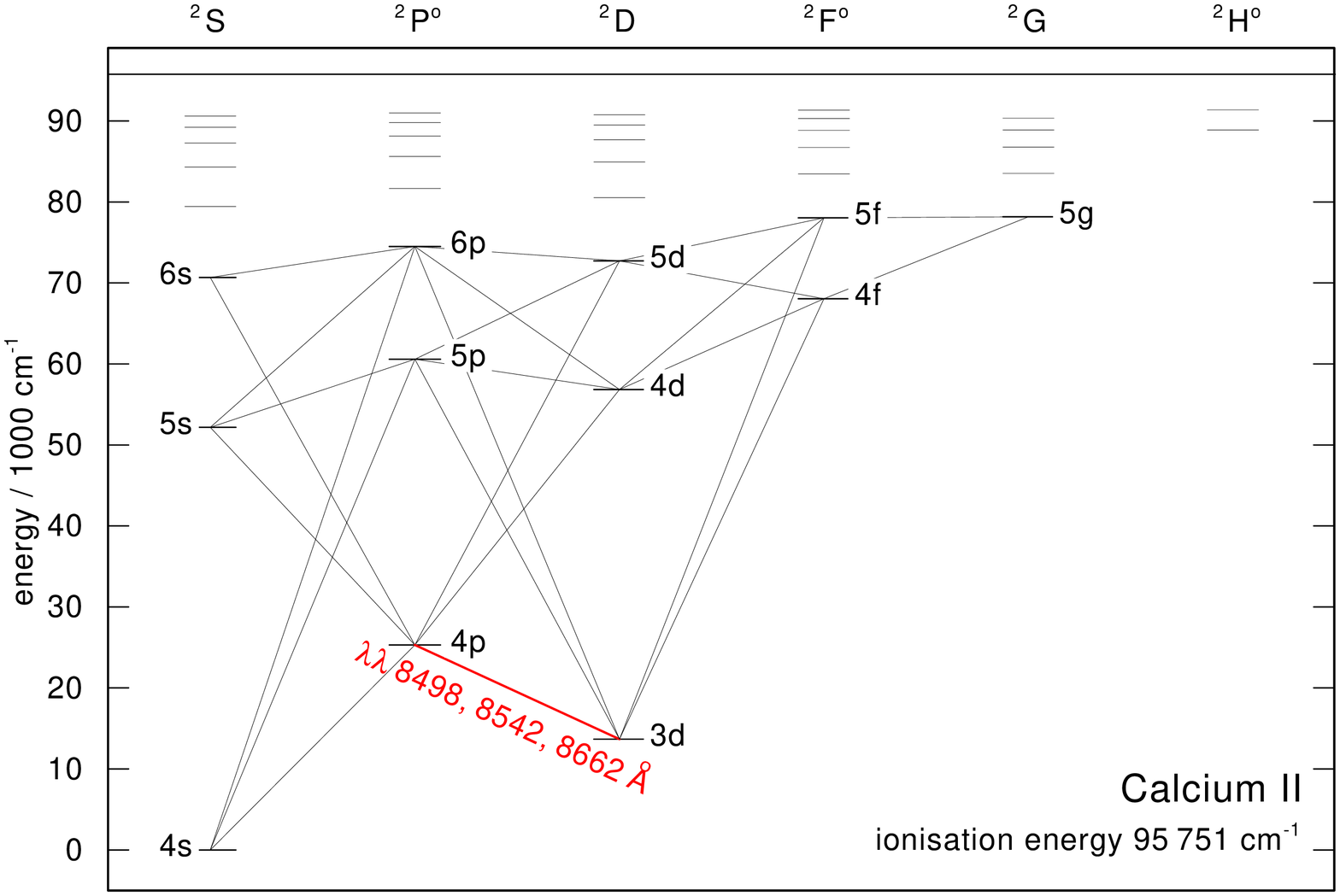}
\end{center}
\caption{
Grotrian diagram of our \ion{Ca}{ii} model ion. All highly excited
levels that are not linked by radiative line transitions are treated in
LTE. The 3d--4p transition causes the observed IR triplet.
}\label{grotrian}
\end{figure}

\section{A viscous disk ring model}

Basically we assume a Shakura \& Sunyaev (1973) $\alpha$-disk model,
i.e., a geometrically thin Keplerian disk heated by viscosity. For
numerical calculations we employ AcDc, our Ac\emph{cretion} D\emph{isk}
c\emph{ode} that we developed for modeling disks in cataclysmic
variables and low-mass X-ray binaries (Nagel \etal 2004). The code
computes a detailed vertical structure and the spectrum of a disk which
is being built from radial-symmetric annuli. For any disk annulus we
assume that it radiates like a plane-parallel slab in non-LTE,
and in radiative and hydrostatic equilibrium. We are presenting here the
results of a single disk ring using different values for the input
parameters \Teff\ (as a measure for the viscously dissipated energy) and
surface mass density $\Sigma$ (the vertical mass column from the disk
midplane to the surface). Further input parameters, which are kept
fixed, are white dwarf's
mass and radius as given above, and $R_{\rm in}$\,=\,91\,R$_{\rm WD}$ and
$R_{\rm out}$\,=\,109\,R$_{\rm WD}$ for the inner and outer disk radii,
respectively. The chemical composition is calcium dominated with an
admixture of hydrogen with varying H/Ca abundance ratios in order to
determine the extent of H-deficiency.

The principal problem for any modeling attempt is posed by the question:
what heats the \ion{Ca}{ii} emission line region? Although being hot, it
cannot be the white dwarf because it is too distant. Also, it cannot be
gravitational energy released through viscosity because the required
mass-accretion rate would be $\approx$\,10$^{-8}$\,\msun/yr which is by many
orders of magnitude larger than the accretion rate invoked for the
presence of settling metals in DAZ photospheres 
($\approx$\,10$^{-15}$\,\msun/yr, Koester \& Wilken 2006). A speculation by Jura
(2008) is additional heating by energy dissipation through disk
asymmetries, being driven by some external unseen planet. As mentioned
above, such an asymmetry is obvious from the line profiles.  Whatever,
at the moment we do not know the heating mechanism. Therefore we need to
use \Teff\ as a free parameter for the disk ring. \emph{In praxi} this
means that we set the accretion rate $\dot{M}$, which is related to
\Teff\ through $$T_{\rm eff}^4(R)=[1-(R_{\rm WD}/R)^{1/2}]\,3GM_{\rm
WD}\dot{M}/8\sigma\pi R^3,$$ to an artificially high value.

\begin{figure}
\begin{center}
\includegraphics[width=\textwidth]{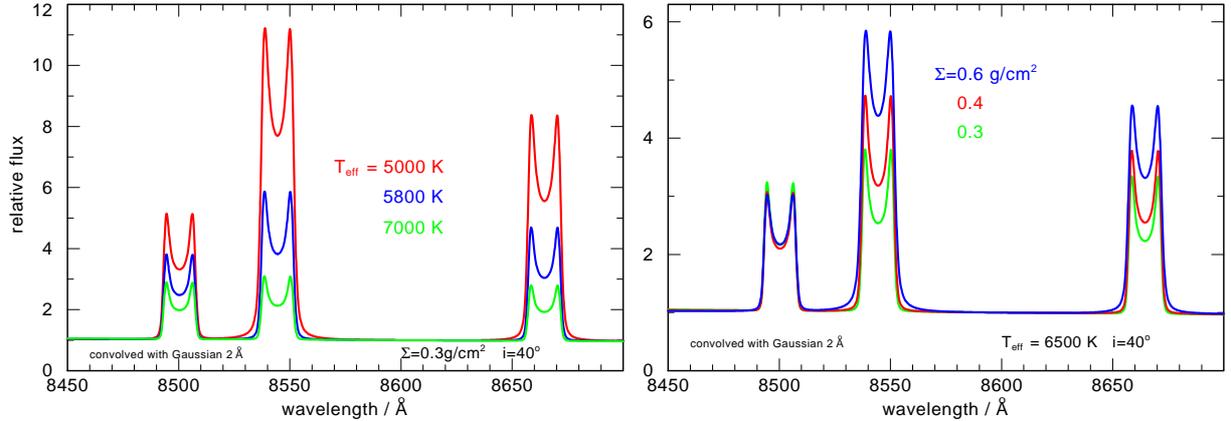}
\end{center}
\caption{\emph{Left:} \Teff\ dependence of the \ion{Ca}{ii}
  triplet emitted by a Keplerian rotating gas ring seen at an
  inclination angle $i$\,=\,40$^\circ$.  The emission becomes weaker with
  increasing \Teff. At the same time, the relative strengths of triplet
  components become equal. \emph{Right:}  A similar effect is seen when
  the surface mass density $\Sigma$ decreases. Values for the ring
  parameters \Teff\ and $\Sigma$ are given in the panels. All profiles
  are convolved with a 2\,\AA\ FWHM Gaussian.}\label{fig_teff_sigma}
\end{figure}

The calcium model atom used for the non-LTE calculations comprises three
ionisation stages, \ion{Ca}{i\,--\,iii} with 7, 12, 1 non-LTE levels and 3,
26, 0 radiative line transitions, respectively. For the disk model
calculations we do not account for level fine structure splitting. For
the \ion{Ca}{ii} IR triplet this is done in the final formal solution of
the transfer equation by distributing the level populations over the fine
structure levels according to their statistical
weight. Fig.\,\ref{grotrian} is a Grotrian diagram of our  \ion{Ca}{ii}
model ion with all implemented levels and lines, indicating the
transition responsible for the observed IR triplet.

Energy levels were taken from {\sc nist}, and oscillator strengths and
photoionisation cross-sections from the Opacity and Iron
Projects. Electron collisional rates are computed from usual
approximation formulae. Rates for collisions with H are currently
neglected, probably without consequences. That is because, first, Mashonkina \etal (2007)
have shown that these collisions are unimportant in F/G type stars
(having similar physical conditions in their photospheres) and,
second, we are faced here with a H-deficient environment. Collisions of
Ca with heavy ions are neglected, too. Their rates are unknown but
probably unimportant because of the low thermal velocities of the
perturbers. For the spectral lines we assume Voigt profiles with
radiation damping parameters. Future improvement should include Van der
Waals damping, although we don't expect it to be important in the disk
environment. The hydrogen model atom is a standard configuration with ten
non-LTE levels that we use routinely for stellar spectral modeling.

\section{Results}

We present first results from disk ring models for \sdss, discussing:

\noindent -- the influence of \Teff, $\Sigma$, and inclination $i$ on emergent line
profiles of the \ion{Ca}{ii} IR triplet;

\noindent -- an upper limit for hydrogen content from H$\alpha$;

\noindent -- the characteristics of one representative model, in particular its vertical
structure;

\noindent -- a comparison of the computed spectrum with observation.

\subsection{Influence of \Teff\, $\Sigma$, $i$ on the infrared \ion{Ca}{ii} triplet}

In Fig.\,\ref{fig_teff_sigma} (left panel) we see that the emission
strength of the \ion{Ca}{ii} triplet decreases with increasing \Teff\
(with $\Sigma$\,=\,0.3\,g/cm$^{2}$ kept fixed). This is because of the
shifting \ion{Ca}{ii}/\ion{Ca}{iii} ionisation balance. A closer comparison
of the three triplet components shows that the relative strengths of the
components become equal with increasing \Teff, a behaviour that
constrains \Teff\ from the observed line strengths. A similar trend is
seen (Fig.\,\ref{fig_teff_sigma}, right panel) when $\Sigma$ is reduced
(with \Teff\,=\,6500\,K kept fixed). We stress that the models have a
considerable continuum flux compared to line emission peak heights.

\begin{figure}
\begin{center}
\includegraphics[width=\textwidth]{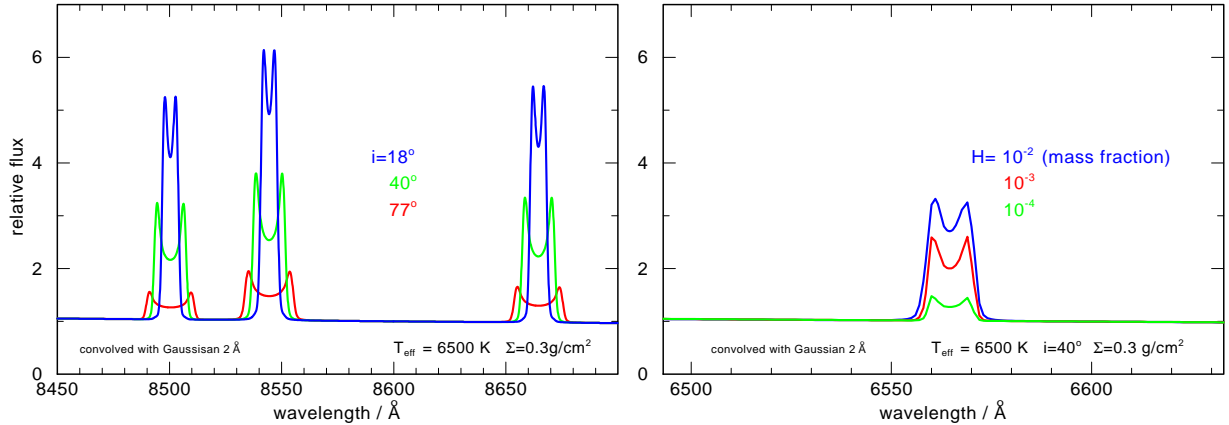}
\end{center}
\caption{\emph{Left:} The \ion{Ca}{ii} line shape of a particular model
  (\Teff\,=\,6500\,K, $\Sigma$\,=\,0.3\,g/cm$^{2}$) seen under different
  inclination angles $i$. The relative strength of the profile
  depression between double peaks increases with $i$. \emph{Right:}
  H$\alpha$ line shape of the same model but with different hydrogen
  content. H\,$>$\,1\% would be detectable in
  \sdss.}\label{inclination_halpha}
\end{figure}

In Fig.\,\ref{inclination_halpha} (left panel) we demonstrate that in
principle the inclination angle can be constrained from the \ion{Ca}{ii}
triplet line shape. Obviously, the relative strength of the profile
depression between the double peaks increases with inclination.

\subsection{Upper limit for hydrogen}

From the lack of H$\alpha$ emission in the spectrum of \sdss\ we can
make a quantitative estimate of the hydrogen-deficiency in the gas disk.
Fig.\,\ref{inclination_halpha} (right panel) depicts the H$\alpha$ line
shape of a particular model in which we varied the H content (H\,=\,1\%,
0.1\%, 0.01\%; mass fraction). With an abundance of 1\% the
H$\alpha$ peak height is comparable to that of the \ion{Ca}{ii} triplet
and, hence, would be detectable in the spectrum of \sdss.

\begin{figure}
\includegraphics[width=7cm]{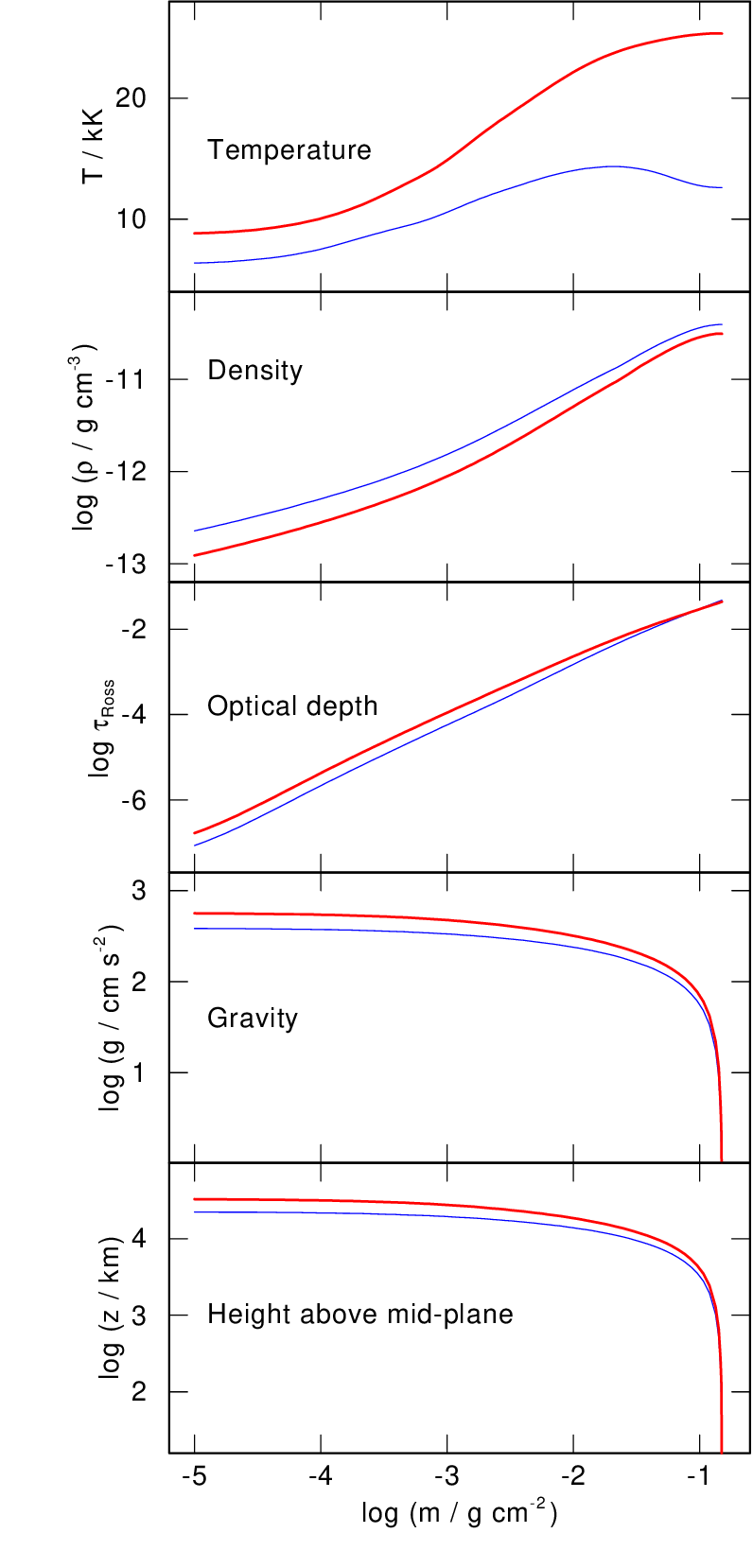}\hspace{2pc}%
\begin{minipage}[b]{8cm}
\caption{{\bf (Left)} Vertical structure of two models
  (\Teff\,=\,5000\,K, thick lines, and 7000\,K)
  with $\Sigma$\,=\,0.3\,g/cm$^{2}$. The run of various physical quantities
  is shown on a column-mass scale measured from the disk surface (left)
  to the disk midplane (right); see discussion in the text
  (Sect.\,\ref{sectvertical}). The disks are optically thin ($\tau_{\rm
  Ross}\,<\,0.1$).\label{vertical_structure}}
 \vspace{5cm}
  \includegraphics[width=7cm]{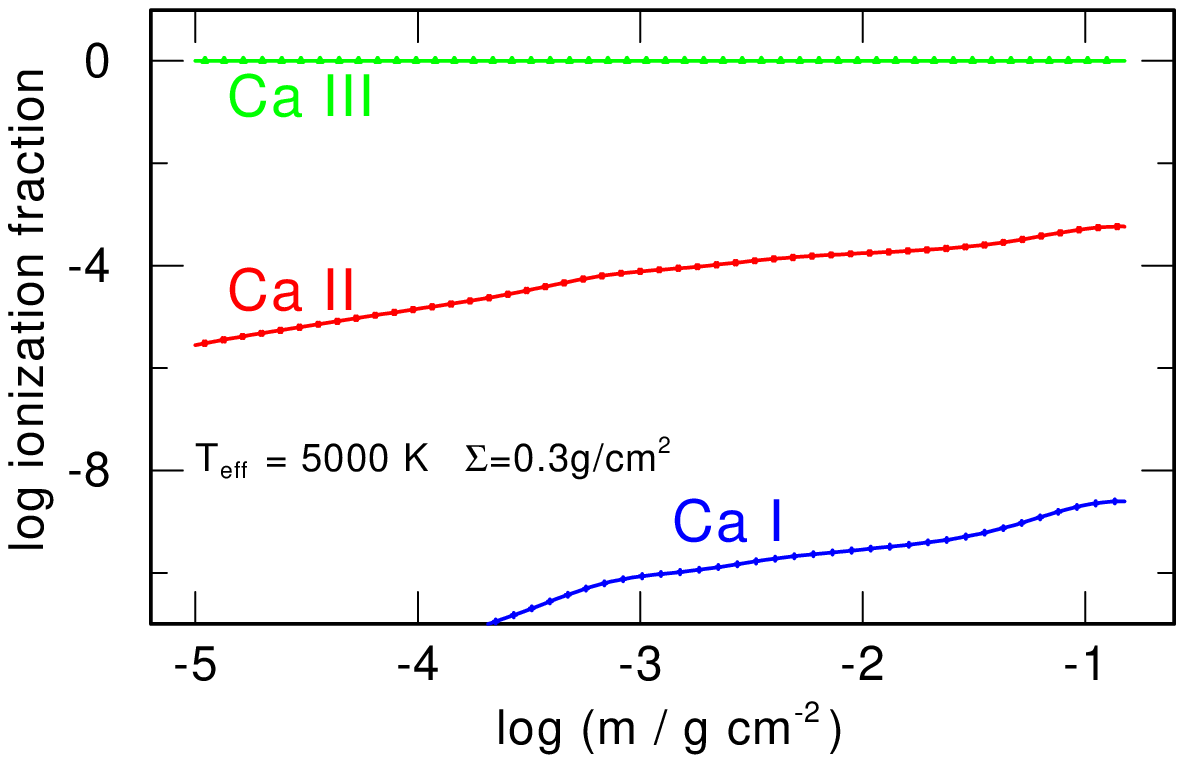}
\caption{{\bf (Above)} Vertical run of the calcium ionisation fractions in the
  model with \Teff\,=\,5000\,K and
  $\Sigma\,=\,0.3$\,g/cm$^{2}$. \ion{Ca}{iii} is the dominant
  ionisation stage.
  \label{ion}}
\end{minipage}
\end{figure}

\subsection{Vertical structure of disk ring}\label{sectvertical}

Let us inspect the vertical structure of two representative ring
models with different effective temperature displayed in
Fig.\,\ref{vertical_structure}. The range of temperature $T$ and mass
density $\rho$ is comparable to the circumstances encountered in the
atmospheres of F/G-type giants. Accordingly, the gravity increases from
the disk midplane toward the disk surface, up to  \logg\,$\approx$\,2.5. The
run of the Rosseland optical depth $\tau_{\rm Ross}$ shows that the
disks are optically thin. The geometrical height $z$ above the midplane
shows that the thickness of the disk ring is about
 $\Delta\,R$\,=\,50\,000\,km, hence, $\Delta R/R$\,$\approx$\,15. The non-LTE departure
coefficients (not
plotted) of the atomic population numbers in these two models deviate
significantly from the LTE value $b$\,=\,1. For the lower levels of the
\ion{Ca}{ii} IR triplet in the \Teff\,=\,5000\,K model, for example, they
range between $b$\,$\approx$\,0.001 in the upper layers of the disk and
$b$\,$\approx$\,50 in deeper regions.

Fig.\,\ref{ion} shows the vertical calcium ionisation stratification for
the cooler (\Teff\,=\,5000\,K) of these two models. \ion{Ca}{iii} is by far
the dominant ionisation stage everywhere in the model. This advises us
to extend the model atom to include the next higher ionisation stage in future
calculations, although we do not expect a dramatic depopulation of
\ion{Ca}{iii} (and \ion{Ca}{ii}) because its ionisation potential is
rather high (50.9\,eV).

\begin{figure}
\begin{center}
\includegraphics[width=\textwidth]{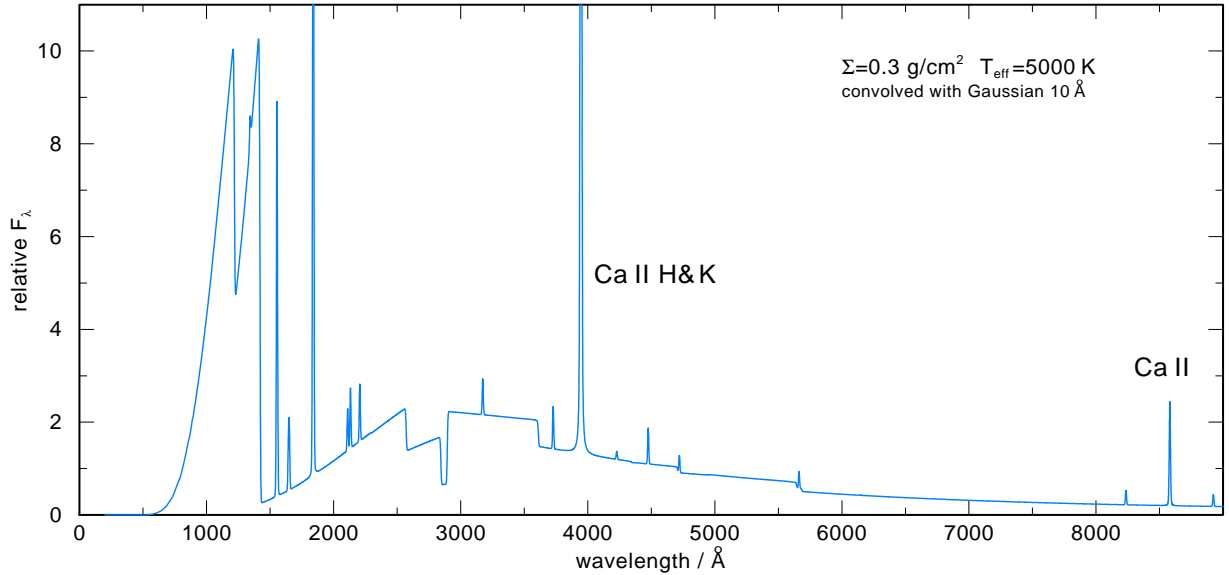}
\end{center}
\caption{Flux distribution in the UV/optical range of the \Teff\,=\,5000\,K,
  $\Sigma$\,=\,0.3\,g/cm$^{2}$ model (no fine structure splitting applied
  to lines). The spectrum is dominated by
  \ion{Ca}{ii} emission lines and bound-free emission edges. The
  strongest feature is the H\&K line. For the sake of clarity, the
  spectrum is convolved with a 10\,\AA\ FWHM Gaussian.}\label{overall_flux}
\end{figure}

\begin{figure}
\begin{center}
\includegraphics[width=9.5cm]{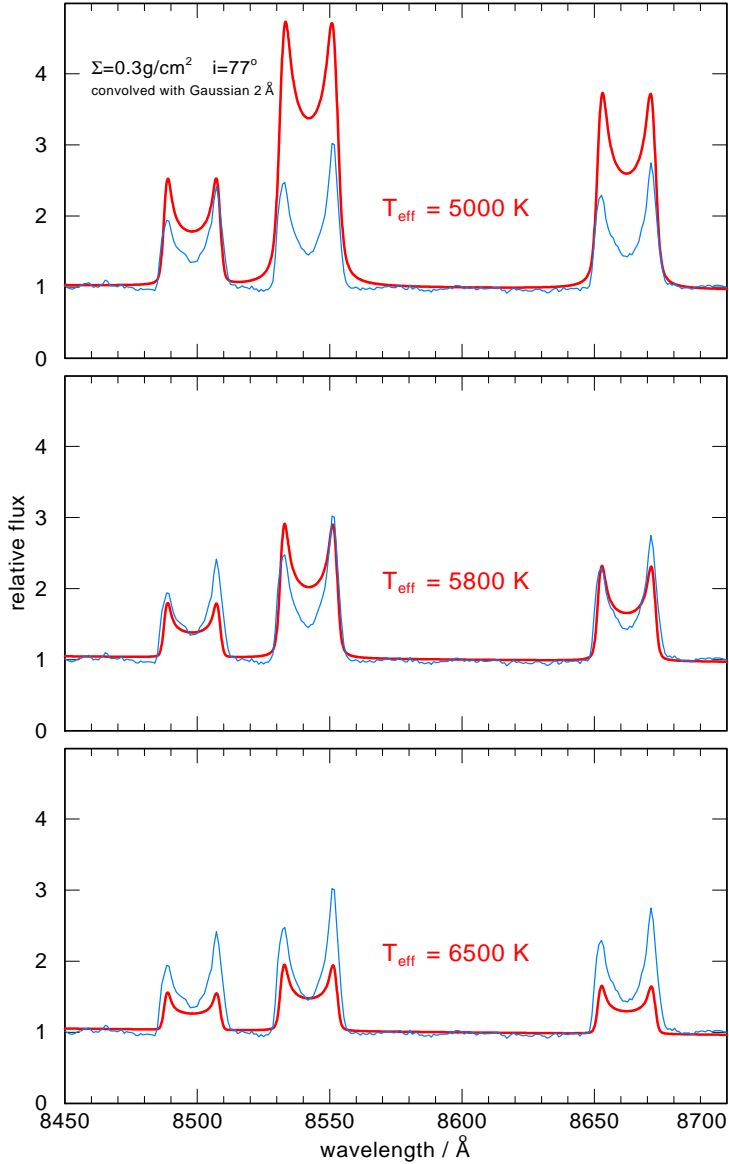}
\end{center}
\caption{Normalized spectra of three models (thick lines) with different \Teff\ compared to
  the observed spectrum of \sdss. The fitting procedure is ambiguous because the observation is
  normalized to the continuum that is probably dominated by the white
  dwarf. All models have $\Sigma$\,=\,0.3\,g/cm$^{2}$ and are viewed at an
  inclination of $i$\,=\,77$^\circ$.}\label{fit}
\end{figure}

It is interesting to look at the overall flux spectrum of the models,
e.g., Fig.\,\ref{overall_flux}. The spectrum is
dominated by \ion{Ca}{ii} emission lines. The \ion{Ca}{ii} H\&K
resonance doublet is by far the strongest emission feature that should be
detectable in the spectrum of \sdss, although the WD flux increases by a
about an order of magnitude from the location of the IR triplet to the
H\&K line. A close inspection of the residual disk spectrum of \sdss\ (i.e.,
observation minus WD model spectrum) presented by G\"ansicke \etal
(2008) indeed displays an emission feature at the location of the H\&K
line, although not explicitly described by the authors.  In a new
high-quality spectrum of \sdss\ (G\"ansicke, this meeting) 
the presence of this emission line is
clearly confirmed. It seems, however, that the observed strength is
strongly overestimated by our models, a result that also holds for the
LTE emission models presented by G\"ansicke.

Another remarkable feature of the overall model flux spectrum in
Fig.\,\ref{overall_flux} is the occurrence of two prominent \ion{Ca}{ii}
\emph{emission edges} at 1219\,\AA\ and 1420\,\AA, created by
recombination processes into the first and second excited states,
respectively. These features should be easily detectable in UV spectra
as well as many other emission lines from other metals as predicted and
presented by G\"ansicke at this meeting. Our models promise that the
proposed (and approved) HST/COS spectroscopic observation of \sdss\ by
G\"ansicke and collaborators will deliver an exciting dataset that will
allow important conclusions on the chemical composition of the gas disk.

\subsection{Models vs. observation}

In Fig.\,\ref{fit} we show a direct comparison of the normalised
spectrum of \sdss\ to three models with different effective
temperatures. This comparison is, however, complicated by the fact that
G\"ansicke \etal (2006) suggest that the continuous radiation in the
vicinity of the \ion{Ca}{ii} triplet stems from the white dwarf alone,
while in contrast our models predict a non-negligible contribution by
the disk. What we compare in Fig.\,\ref{fit} is the observed spectrum
normalized to the WD (+disk) continuum and the computed disk spectrum
alone, normalized to its emission continuum. In any case, it can be seen
that the effective temperature of the disk at \sdss\ is well constrained by the
three models, being \Teff\,$\approx$\,5800\,K. The cooler model
(\Teff\,=\,5000\,K) is perhaps more favorable because of the larger
line-to-continuum emission ratio, while the hotter model (\Teff\,=\,6500\,K)
has the advantage that the relative strengths of the three line
components are matched better.

\section{Summary and conclusions}

The infrared \ion{Ca}{ii} emission triplet in the spectrum of the DAZ
white dwarf \sdss\ can be modeled with a geometrically and optically
thin, Keplerian  viscous gas disk ring at a distance of 1.2~$\rsun$ from
the WD, with \Teff\,$\approx$\,6000\,K and a low surface mass density 
$\Sigma$\,$\approx$\,0.3\,g/cm$^{2}$. One serious open problem is the
unknown 
disk-heating mechanism. The disk is hydrogen-deficient (H\,$\leq$\,1\% by mass)
and it is located within the tidal disruption radius ($R_{\rm
tidal}$\,=1.5\,\rsun). If one assumes that the disk reaches down to the WD
and that it has uniformly this surface density, then its total mass
would be 7$\cdot\,10^{21}$\,g. A rocky asteroid
($\bar{\rho}$\,=\,3\,g/cm$^{3}$) with this mass would have a diameter of
about 160\,km. An asteroid of this size in our solar system is, e.g., 22~Kalliope.

Future work will include other abundant metal species, with a
composition appropriate for asteroids. However, at the moment there are
only two \ion{Fe}{ii} lines detected. Real progress for an abundance
analysis of the gas disk around \sdss\ is only possible with future UV
spectroscopy.

While the determination of the composition of accreted material from
abundance analyses in DAZ atmospheres is an indirect method that depends
on our theoretical knowledge about metal settling timescales, the
analysis of the gas disks has the obvious advantage that is a direct
composition measurement.

\ack We thank Boris G\"ansicke for sending us his \sdss\ spectrum in
electronic form and for useful discussions. T.R. is supported by the
\emph{German Astrophysical Virtual Observatory} project of the Federal
Ministry of Education and Research (grant 05\,AC6VTB).

\section*{References}
\begin{thereferences}

\item Becklin, E. E., Farihi, J., Jura, M., Song, I., Weinberger,
  A. J., \& Zuckerman, B. 2005, ApJ, 632, L119

\item Debes, J. H., \& Sigurdsson, S. 2002, ApJ, 572, 556

\item G\"ansicke, B. T., Marsh, T. M., Southworth, J., \&
  Rebassa-Mansergas, A. 2006, Science, 314, 1908

\item G\"ansicke, B. T., Marsh, T. M., \& Southworth, J. 2007, MNRAS,
  380, L35

\item G\"ansicke, B. T., Marsh, T. M., Southworth, J., \& Rebassa-Mansergas, A.  2008,
  preprint, arXiv:0710.2807v1

\item Graham, J. R., Matthews, K., Neugebauer, G., \& Soifer, B. T. 1990,
  AJ, 357, 216

\item Jura, M. 2003, ApJ, 584, L91

\item Jura, M., Farihi, J., \& Zuckerman, B. 2007, ApJ, 663, 1285

\item Jura, M. 2008, AJ, 135, 1785

\item Kilic, M., von Hippel, T., Leggett, S. K., \& Winget, D. E. 2005,
  ApJ, 632, L115

\item Kilic, M., von Hippel, T., Leggett, S. K., \& Winget, D. E. 2006,
  ApJ, 646, 474

\item Koester, D., \& Wilken, D. 2006, A\&A, 453, 1051

\item Koester, D., Provencal, J., \& Shipman, H. L. 1997, A\&A, 320, L57

\item Mashonkina, L., Korn, A. J., \& Przybilla, N. 2007, A\&A, 461, 261

\item Nagel, T., Dreizler, S., Rauch, T., \& Werner, K. 2004, A\&A, 428, 109

\item Reach, W. T., Kuchner, M. J., von Hippel, T., \etal 2005, ApJ, 635, L161

\item Shakura N.~I., \& Sunyaev, R.~A. 1973, A\&A, 24, 337

\item Zuckerman, B., \& Becklin, E. E. 1987, Nature, 330, 138

\end{thereferences}

\end{document}